\documentclass[preprintnumbers, floatfix, preprintnumbers, letterpaper, superscriptaddress,nofootinbib, twocolumn]{revtex4}
\pdfoutput=1
\usepackage{graphicx}
\usepackage{microtype}
\usepackage{amsmath}
\usepackage{amssymb}
\usepackage{subfigure}
\usepackage{hyperref}
\usepackage{url}
\usepackage{xcolor}
\usepackage{color}
\usepackage{mathrsfs}
\usepackage{calrsfs}
\usepackage{amsfonts}
\usepackage{latexsym}
\usepackage{ragged2e}
\usepackage{epsfig}
\usepackage{textcomp}
\usepackage{phaistos}
\usepackage[utf8]{inputenc}
\makeatletter
\renewcommand\@makefnmark{\hbox{\@textsuperscript{\normalfont\color{purple}\@thefnmark}}}
\renewcommand\@makefntext[1]{%
  \parindent 1em\noindent
            \hb@xt@1.8em{%
                \hss\@textsuperscript{\normalfont\@thefnmark}}#1}
\makeatother

\usepackage{caption}
\DeclareCaptionJustification{justified}{\leftskip=0pt \rightskip=0pt \parfillskip=0pt plus 1fil}
\definecolor{vividviolet}{rgb}{0.62, 0.0, 1.0}
\definecolor{amaranth}{rgb}{0.9, 0.17, 0.31}
\definecolor{palatinateblue}{rgb}{0.15, 0.23, 0.89}
\definecolor{brightpink}{rgb}{1.0, 0.0, 0.5}
\definecolor{cornflowerblue}{rgb}{0.39, 0.58, 0.93}
\definecolor{deepcarminepink}{rgb}{0.94, 0.19, 0.22}
\definecolor{radicalred}{rgb}{1.0, 0.21, 0.37}

\hypersetup{ linktoc=all,
    colorlinks, linkcolor={palatinateblue},
    citecolor={brightpink}, urlcolor={amaranth}
}
\newcommand{\changeurlcolor}[1]{\hypersetup{urlcolor=#1}}
\graphicspath{{Images/}}

\renewcommand{\d}[1]{\ensuremath{\operatorname{d}\!{#1}}}

\graphicspath{{Images/}}
\renewcommand{\d}[1]{\ensuremath{\operatorname{d}\!{#1}}}
\makeatletter
\def\@fnsymbol#1{\ensuremath{\ifcase#1\or $\textleaf$ \or $\PHplaneTree$
\else\@ctrerr\fi}}%

\makeatother

\def\sideremark#1{\ifvmode\leavevmode\fi\vadjust{\vbox to0pt{\vss
 \hbox to 0pt{\hskip\hsize\hskip1em
 \vbox{\hsize1.5cm\tiny\raggedright\pretolerance10000
 \noindent #1\hfill}\hss}\vbox to8pt{\vfil}\vss}}}%
\begin{document}

\title{Spacetime Singularities and Cosmic Censorship Conjecture: \\A Review with Some Thoughts}

\author{Yen Chin \surname{Ong}}
\email{ycong@yzu.edu.cn}
\affiliation{Center for Gravitation and Cosmology, College of Physical Science and Technology, Yangzhou University, \\180 Siwangting Road, Yangzhou City, Jiangsu Province  225002, China}

\begin{abstract}
The singularity theorems of Hawking and Penrose tell us that singularities are common place in general relativity. Singularities not only occur at the beginning of the Universe at the Big Bang, but also in complete gravitational collapses that result in the formation of black holes. If singularities -- except the one at the Big Bang -- ever become ``naked'', i.e., not shrouded by black hole horizons, then it is expected that problems would arise and render general relativity indeterministic. For this reason, Penrose proposed the cosmic censorship conjecture, which states that singularities should never be naked. Various counterexamples to the conjecture have since been discovered, but it is still not clear under which kind of physical processes one can expect violation of the conjecture. In this short review, I briefly examine some progresses in spacetime singularities and cosmic censorship conjecture. In particular, I shall discuss \emph{why} we should still care about the conjecture, and whether we should be worried about some of the counterexamples. This is not meant to be a comprehensive review, but rather to give an introduction to the subject, which has recently seen an increase of interest.
\end{abstract}

\maketitle

\section{Introduction: Are Spacetime Singularities Verboten?}\label{1}

Singularities are one of those peculiarities in general relativity that are of interests not only to experts but also to the general public. Admittedly, the concept of black hole is already strange enough. Theoretical physicists who work with black hole solutions on a daily basis have perhaps gotten used to their various properties, but if we stop and think for a moment, and really appreciate that some of these are \emph{vacuum} solutions of the Einstein field equations, it is quite remarkable that \emph{nothingness}\footnote{It is quite misleading to say that the ``central singularity'' is the ``source'' of the mass. See the subsequent discussion for details.}, ``just'' empty spacetime, can bend light rays and affect particle dynamics, and even rotate and drag stuff around it (the Kerr ergoregion), not to mention preventing anything from escaping an event horizon that is not even a physical barrier but a mathematical boundary. 

Despite all this remarkable features of black holes, singularities that stay hidden inside them are even more mysterious, as we hardly know anything about their properties. It is often said that general relativity predicts its own demise, since singularities are predicted by general relativity and yet cannot be described by the theory. 
This, in and by itself, is not much of a concern, since it is widely believed that a quantum theory of gravity will eventually resolve the singularity. In this point of view, a singularity is an \emph{artifact} of applying general relativity in the regime that it cannot be trusted. From a modern effective field theory perspective (see e.g., \cite{9512024,0311082}), there is a cutoff scale below which an effective field theory breaks down, and new degrees of freedom would enter the physics that need to be taken into considerations. In the case of general relativity, when curvature becomes extremely large near the singularities, one would expect quantum effects to be important. Therefore, it is hoped that once this is properly taken into account, the singularities would disappear. In addition, since these singularities are hidden inside black holes and therefore cannot influence the exterior universe, the conventional wisdom is that they can be safely ignored. Out of sight is out of mind. 

The problem is this: what if, under some physical processes, the singularity inside a black hole becomes naked, i.e., the horizon is destroyed and the singularity is revealed? As we shall soon discuss, this will indeed be problematic. For a long time it is hoped that no physical process can create a naked singularity. Eventually some counterexamples are found and then the best hope becomes: for \emph{generic}
initial conditions, naked singularity cannot be formed. This is the ``Cosmic Censorship Conjecture'' \cite{1,1-2}. The conjecture is further refined into the strong and weak version, as we shall see. Until now there is no good understanding of what ``generic'' really means in this context, as more and more counterexamples involving a variety of physical processes have been found. 

This short review is not meant to be comprehensive. Its main aim is to serve as an introduction to the subject, which has seen a remarkable increase in interest in the recent years (one need only checks the citation history of Penrose's 1965 paper \cite{1} on inSPIRE to see this). 
It contains only three parts: in Sec.(\ref{sec2}), I shall first explain what spacetime singularities are (there are many confusions even about the simplest case of Schwarzschild black hole) and why we should care about them, especially when they become naked, but \emph{also} when they are hidden inside black holes. In Sec.(\ref{sec3}) I will review some counterexamples to cosmic censorship conjecture, with focus on the recent progress in this subject. In addition, I will also review some approaches that attempted to break cosmic censorship but nevertheless the censorship holds. In Sec.(\ref{sec4}) I will discuss the implications of the counterexamples, i.e. whether censorship can be saved in some way or another, and whether we should worry about naked singularities. I will also discuss some curiosities such as the possible connection to weak gravity conjecture. Lastly I will discuss some prospects for future investigations in this subject, which is of course somewhat biased due to my limited involvements in this field.
Indeed, the entire review would no doubt be affected by my personal experience, thoughts, and prejudices, which I hope would nevertheless not affect most of the objectiveness of the sciences involved.

\section{What Are Singularities and Why Should We Care?}\label{sec2}

There are two major types of spacetime singularities.  
The first type is what we usually have in mind when we, as physicists, mention spacetime singularities: the points or regions where curvature becomes infinite. Since the components of the Riemann curvature tensor is coordinate dependent, one often constructs scalar curvature invariants for this purpose. When we say that curvature is divergent, what we mean implicitly is that the scalar curvature invariants, such as the Ricci scalar $R:=g^{\mu\nu}R_{\mu\nu}$, and the Kretschmann scalar $K:=R^{\mu\nu\lambda\rho}R_{\mu\nu\lambda\rho}$, diverge. Note that strictly speaking, curvature singularity is not considered a part of the spacetime manifold. \emph{In practice}, however, singularity is understood as location at which the curvature is \emph{arbitrarily large}, but need not be actually infinite. The blow-up rates of curvature during gravitational collapse can be studied with great mathematical details \cite{2003.13330, 2004.11831}.
 
The second type is in the sense of geodesic incompleteness of the spacetime manifold. That is, the worldline of some particle cannot be extended beyond a certain proper time or affine parameter. In other words, singularities are \emph{obstructions} to extend causal geodesics. This definition has nothing to say about the curvature, at least \emph{a priori}.
It is this notion of singularity that is discussed in the singularity theorem of Penrose \cite{1}, which states: if a spacetime contains a non-compact Cauchy hypersurface $\Sigma$ (``causality condition'') and a closed future-trapped surface $\mathcal{H}$ (``trapping condition''), and if the null Ricci condition $R_{\mu\nu}k^\mu k^\nu \geqslant 0$ holds for all null vector $k$ (``energy conditions'), then there exist future incomplete null geodesics. In other words, light rays hit a singularity in finite affine parameter. As a side remark, the null Ricci condition is better known in the literature and textbooks as the ``null energy condition'' (NEC), but it is really a geometric condition. See \cite{jose}, and also \cite{parikh}. In this short review, we will not be too careful with rigorous mathematical definitions, a good source for that for interested readers is the old review in 1999 by Penrose himself \cite{penrose1999}.

However, these two notions of singularity usually coincide in the case of black hole solutions in general relativity (see also Section 5.1.5 of \cite{jose} and the references therein). The reason is this: if there exists incomplete timelike or null geodesics \emph{mathematically}, then we should ask where and why does the geodesic end \emph{physically}. For example, one can remove by hand one point in Minkowski spacetime and it would be geodesically incomplete, but this is not very physical (indeed, mathematically it can be easily patched up again -- it is an example of what is known an extendible singularity). On the contrary, it is arguably physically natural that geodesics end when something very drastic happens to the geometry, such as when curvature becomes divergent. Note that in the singularity theorem the metric is assumed to be ``sufficiently nice''. Technically this means that the metric has to be at least twice differentiable, i.e., of class $C^2$. A generalization of the theorem to metric tensors of class $C^{1,1}$, i.e., the first derivative is Lipschitz continuous, has been achieved rather recently \cite{1502.00287}. 

The implication of Penrose's theorem is huge: it means that singularities are ubiquitous in general relativity. Since black holes have been observed to be common occurrences in the actual Universe, this also means that singularities are common features of the Universe. Singularities are therefore not only a curiosity of general relativity, but an essential \emph{prediction} of the theory. Nevertheless, as we briefly discussed in the Introduction, it is commonly believed that singularities can be resolved by quantum gravity, and therefore we can safely ignore them. I would argue that this is not necessarily the case, and even if it is true, there are still some subtleties that we have to take care of. The devil is in the detail.

First of all, since we do not yet have a full working theory of quantum gravity, there is simply \emph{no} guarantee that quantum effects, when properly taken into account -- will resolve spacetime singularity. 
There is in fact one lesser known possibility: quantum effects might be \emph{suppressed} near spacetime singularities. For example, the string coupling of the dilatonic charge GHS black hole \cite{ghs,g,gm} becomes weak at the singularity \cite{9210119}, suggesting that it is behaving ``classically'', whatever that might mean. Of course, just like in general relativity, the GHS solution is still a low energy limit of string theory, so it is not to be trusted at the singularity. Consequently, this behavior is only suggestive at best. There are other possibilities that are consistent with weak quantum effect at the singularity, ranging from treating $\hbar$ as a field whose role becomes increasing small at higher energy \cite{1208.5874,1212.0454}, to asymptotic safe gravity \cite{0610018} (holding the Planck mass fixed in 4-dimensions, varying Newton's constant $G \to 0$ is equivalent to varying $\hbar \to 0$) and various considerations from generalized uncertainty principle \cite{0912.2253,1504.07637,1804.05176,1806.03691,1905.00287}. 

Surprisingly, it was also argued that if we do resolve singularities in string theory, this could give rise to other problems. 
For example, a Schwarzschild-AdS (Anti-de Sitter) metric with a negative mass corresponds to a naked timelike singularity. Once this singularity is resolved, there would appear states that possess arbitrarily negative energy. By holography this means that the corresponding dual field theory on the boundary has a Hamiltonian that is not bounded below and therefore unphysical \cite{9503062, 0410040}. Note that this arbitrary negative energy states in the AdS bulk does not violate the AdS version of positive energy theorem because holography is a duality between a \emph{stringy} bulk and quantum field theory on the boundary, i.e. the effective theory in the bulk actually involves higher order terms that spoils the conditions for the energy theorem\footnote{Often in practice, e.g., in AdS/QGP (quark-gluon plasma) correspondence, one just uses general relativity solutions in the bulk. This is valid in modeling some phenomena but certainly \emph{not} when dealing with naked singularity -- higher order terms cannot be neglected in the presence of large curvature.}. The same argument -- minus the holography part -- of course also applies to asymptotically flat black holes \cite{9503062}. In other words, even \emph{if} we believe that quantum gravity can resolve singularities, there are subtleties that we do not quite understand. Indeed, even without knowing the fiull details of how singularities might be resolved, many so-called ``regular black holes'' without singularities have been proposed and widely studied in the literature. The first \emph{exact} solution was provided by a charged black hole supported by nonlinear electrodynamics \cite{9911046}, which is actually a charged version of  
Bardeen black hole \cite{bardeen}. A recent study, however, shows that the set of regular geometries that arises is remarkably limited, and furthermore getting rid of the singularities also seem to lead to significant modification \emph{outside} the black hole \cite{1911.11200}; see also \cite{1510.04957}. These solutions are also known to be unstable, or not fully predictive regarding the final moments of Hawking evaporation \cite{1805.02675}.

To summarize: we cannot be sure that quantum gravity can resolve singularities, even if it does, we do not quite know how it does so without giving rise to other problems. In addition, anything that falls into a Schwarzschild black hole would crash into the singularity in a finite proper time of at most $\pi M$, where $M$ is the mass of the black hole\footnote{In this review, unless otherwise specified, we use the so-called relativistic units \cite{werner} such that Newton's constant and speed of light is unity ($G=c=1$), and vacuum permittivity satisfies $4\pi \epsilon_0=1$. We keep the reduced Planck constant $\hbar$ explicit. Thus mass and time has dimension of length, and $\hbar$ has the dimension of area.}. Perhaps the full resolution of the information paradox, which concerns what happens to the quantum information that falls into a black hole (whose conservation is required by unitarity but in tension with the thermal property of the Hawking radiation) would also require a full understanding of the singularity. One proposal involves imposing a \emph{final} boundary condition at the singularity to teleport the information back out \cite{0310281,0311269,0806.3818}. Therefore, regardless of whether singularities can become naked, it is important that we look deeper into the properties of singularities. This is of course, easier said than done.

In view of the above discussions that we cannot simply hope that quantum gravity would resolve singularity, it makes sense to worry about violations of cosmic censorship conjecture. It is now appropriate to further understand the conjecture, which comes in two versions. The \emph{weak} cosmic censorship conjecture is simply the statement that singularities should not be naked. The \emph{strong} cosmic censorship requires that ``physically relevant'' spacetimes should be globally hyperbolic, i.e. if we start with initial conditions on a specified spacelike hypersurface $\Sigma$ at some time $t=t_0$, any spacetime event $p$ lying in the causal future of $\Sigma$ should be fully determined by information on $\Sigma$. These two versions of the conjecture are actually quite similar, but it is easier to understand the weak version first. To this end, let us first consider a rather unphysical spacetime that nevertheless has some important features that would be helpful to our understanding.

Consider a negative mass Schwarzschild spacetime, whose metric tensor is given by simply allowing the mass $M$ of a normal Schwarzschild black hole spacetime to be negative: 
\begin{equation}
\d s^2 = -\left(1+\frac{2|M|}{r}\right) \d t^2 + \left(1+\frac{2|M|}{r}\right)^{-1} \d r^2 + r^2 \d \Omega^2,
\end{equation}
where $\d \Omega^2$ is the standard round metric on a sphere. This spacetime is not physical due to the positive energy theorem.

Clearly there is no event horizon in this spacetime, and $r=0$ is a \emph{timelike} singularity. Consider any regular point $p$ in this spacetime. Its causal past would intersect with the timelike singularity, which means that information from the singularity would affect $p$. In other words, we must impose some boundary conditions on the singularity in order to determine what happens at any point in the spacetime. This we cannot do, however, because we do not know the physics of singularities. Therefore, determinism is completely spoiled if such a singularity is formed. Thus, the problem of singularity lies not only with itself and its surrounding spacetime of high curvature region, but also with its \emph{influence} on spacetime region far away. 

The weak cosmic censorship conjecture is fine with singularities inside black holes that are timelike, such as the ones inside charged or rotating black holes, since by the very definition of black hole, information about the singularities cannot affect the external world. 
The strong cosmic censorship, on the other hand, essentially requires that the future should be fully deterministic even inside a black hole. After all, an observer can be sent into an arbitrarily large black hole, and if he or she survives the trip, could perform physical observations and experiments inside of the black hole. At least in principle then, one can do physics inside black holes, just not report the findings back to colleagues in the exterior spacetime. Since we should not discriminate against these brave observers, physics had better be deterministic regardless of whether one is inside a black hole or not. 

In charged or rotating black holes in general relativity, there are two horizons. If we take these geometries at face value, then after the observer falls through the inner horizon, he or she can be influenced by the singularity. The problem is the same as the one with naked singularity discussed before. In fact, the singularity inside charged or rotating black holes is ``naked'' with respect to the observer inside the inner horizon. It is not difficult to show that the ``effective mass'' of such a ``naked singularity'' is also negative\footnote{An easy way to appreciate this is to consider the weak field approximation in general relativity, which relates the Newtonian potential $\Phi$ and the metric component $g_{tt}$. In the Schwarzschild black hole case, we have \begin{equation}
1-\frac{2M}{r}= 1+2\Phi.
\end{equation}
That is, $\Phi=-M/r$, which gives the Newtonian force $F=-\nabla \Phi=-M/r^2$. In the Reissner-Nordstr\"om case, we would have instead,
\begin{equation}
1-\frac{2M}{r} + \frac{Q^2}{r^2}= 1+2\Phi,
\end{equation}
so that $\Phi=-M/r+{Q^2}/{2r^2}$. Thus the potential is zero at 
\begin{equation}
{r}=\frac{Q^2}{2M}.
\end{equation}
The \emph{force} is zero at 
\begin{equation}
{r}=\frac{Q^2}{M}.
\end{equation}
Near the vicinity of the singularity below this value of $r$, gravity becomes repulsive (see also, \cite{1401.0741}). A more rigorous approach can be made with quasi-local energy, to show that the mass-energy of the singularity is negative (see, e.g. the review paper \cite{sza}), despite the fact that the entire ADM mass of the spacetime is positive.}, much like the negative mass Schwarzschild singularity. Observers are not doomed to hit the timelike singularities in Reissner-Nordstr\"om and Kerr black holes. In fact, due to the repulsive gravity near the singularity, hitting it is not so easy a task.

In mathematics, if an initial hypersurface uniquely determines the events in its causal future, we said that the \emph{Cauchy problem} is well-posed. Since this is no longer true beyond the inner horizon, the inner horizon is also called the Cauchy horizon. 
In order to avoid the problem of indeterminism, the strong cosmic censorship conjecture requires that the Cauchy horizon be unstable against perturbations. In other words, the geometries below the Cauchy horizon described by the Reissner-Nordstr\"om and Kerr metrics are not physically realistic. It is expected that perturbation from matter fields and radiation would cause the unstable Cauchy horizon to be destroyed and the singularity within becomes spacelike. 
This would resolve the problem (but what \emph{is} the metric of a physically realistic charged or rotating black hole?) Progresses to verify this scenario have been made very recently: in \cite{1902.08323, 1905.04613, 2001.02788}, the authors showed that in both Reissner-Nordstr\"om and Kerr spacetimes coupled to a scalar field, the Cauchy  horizon develops a null singularity and the central singularity becomes spacelike. On the other hand, holographic argument suggests that black holes in anti-de Sitter spacetimes respect strong cosmic censorship, with the exception of rotating BTZ black hole \cite{1911.12413}, which supports an earlier study \cite{1906.08265}.

I shall take this opportunity to clarify the importance of the distinction between timelike and spacelike singularities.
Not all singularities are made equal! Every student of general relativity course learns early on that Schwarzschild black holes harbor a singularity, but not all of them appreciate the fact that it is a singularity \emph{in time} instead of a location at the non-existent ``center'' of the black hole. In fact this misunderstanding persists into many research literature. This is not a complaint about semantics; the difference between a timelike singularity and a spacelike one makes all the difference in general relativity when it comes to issues about naked singularities and cosmic censorship. 

A spacelike singularity such as the one inside a Schwarzschild black hole is not a problem because it lies in the future of the observers who fall through the event horizon. Any observer inside cannot see the singularity, one does not fall towards the Schwarzschild singularity like how one falls into a cavern, but rather falls \emph{forward in time}. The singularity is inevitable just like the next Monday will certainly arrive; it is not ``reached'', but rather it ``happened''. No information can propagate from the singularity towards the observer (which amounts to travel backwards in time). 
Note also that since the singularity lies in the future, it is misleading to think of the singularity as sourcing the mass $M$ of the black hole spacetime, although one could of course formally give it a Green function construction much like in classical Newtonian Poisson equation \cite{mcinnes}. 

Even the Big Bang singularity at the beginning of the Universe, which is the subject of Hawking-Penrose singularity theorem \cite{HawkingPenrose}, does not violate the cosmic censorship conjecture\footnote{In modern cosmology, it is widely believed that in the very early Universe there is a phase of exponentially accelerating expansion known as the inflation. One might question whether some kind of singularity theorem still holds under the inflationary paradigm. Indeed it does, and the result is known as the Borde-Guth-Vilenkin theorem \cite{0110012}.}. It is true that we expect quantum gravity effects to be strong at the Big Bang, and so the physics there is not well understood. However, since it is a spacelike singularity that lies in our past, we can take an initial hypersurface some time after the Big Bang, during which physics is already well-understood, and the evolution afterward would be fully determined by Einstein's equations (and particle physics, thermodynamics etc.) Our ignorance of quantum gravity does not affect our ability to calculate evolution of physical systems in this case. The same cannot be said of timelike singularities like that in negative mass Schwarzschild solution, or those inside charged or rotating black holes. The distinction between timelike and spacelike singularities are therefore crucial. In the strong cosmic censorship conjecture, the hope is that the Cauchy horizon is unstable, which in turn would turn the singularity to a spacelike one and therefore harmless (while negative mass Schwarzschild spacetime would be forbidden by the weak cosmic censorship conjecture -- and at least classically by the positive energy theorem). 

The mechanism of this instability is well-known, it is called ``blueshift instability'' or ``mass inflation''. It is well known that viewed from outside, an object that falls into the black hole appears to slow down (gravitational time dilation) and it appears redder, eventually its image darkens so much one can no longer see it. Essentially, this \emph{gravitational redshift} happens as light loses energy while climbing out from the gravitational well around black holes. In the case of inner horizon of a charged or rotating black hole, incoming particle has their energy \emph{blueshifted} on the horizon. This huge amount of energy caused a backreaction on spacetime geometry, which is generally believed to destroy the inner horizon. 

It is somewhat ironic that the harmless spacelike Schwarzschild singularity turns into a timelike one no matter how small a charge is absorbed by the black hole, but then no matter how small the charge, the blueshift instability is supposed to de-stabilize the inner horizon and turns the singularity back into a spacelike one. 

Having explained the weak and strong cosmic censorships and the mechanism involved in the latter, let us now discuss the various attempts 
to violate cosmic censorship, and which of these attempts are successful. 

\section{Examples and Counterexamples to Cosmic Censorship}\label{sec3}

In 1993, Choptuik \cite{choptuik} showed that under gravitational collapse of massless scalar field, there exists a critical parameter $p_c$ (a parameter $p$ governs the strength of the interaction) below which there is no black hole formation, but above which a black hole forms. Near the critical value $p_c$, the mass of the black hole is arbitrarily small, and has the form $M \sim (p-p_c)^\gamma$, where $p > p_c$ and $\gamma \approx 0.37$ is a universal constant for all families of such solutions. As $p \to p_c$ we have a black hole with vanishing mass and a naked singularity forms. However, such a naked singularity is obtained as a specific solution in the one parameter family, i.e.,  very special initial conditions are required to form such a naked singularity. 

The weak cosmic censorship conjecture now precludes such examples as genuine counterexamples, requiring only that \emph{generic} initial conditions do not form naked singularities. Nevertheless, massless scalar field collapse plays very important roles in the early development of cosmic censorship research. Although Nature does not seem to admit massless scalar field, but such a field is very much ``physical'' in the sense that its equation of motion satisfies  quasilinear second order hyperbolic PDE (i.e. has healthy propagation properties). Why should cosmic censorship care (or know) whether a field is realized in the \emph{actual} Universe, as long as it is logically consistent? 
Precluding these counterexamples on the ground that they are not realized in Nature is therefore not a very strong argument, and the formulation of the conjecture in terms of ``genericity'' became more acceptable.

It is also worth mentioning that in 1994, Christodoulou rigorously proved that naked singularity \emph{can} form under gravitational collapse of massless scalar field in general relativity \cite{christodoulou}, though as argued in his later paper, such a singularity is unstable \cite{christodoulou2} and therefore does not constitute a violation of the cosmic censorship. See also \cite{1710.02922} for a recent improved, more robust, proof. There is, by now, a huge literature about naked singularities formed under gravitational collapse, see, e.g. the monograph by Joshi \cite{joshi}, and the reviews \cite{9805066, 0711.4620}. In some works, the initial data set that leads to the formation of naked singularities are not of measure zero. For a recent example, see \cite{2001.04367}.

Another way to try to create naked singularities is to \emph{destroy a black hole} (or more precisely, its horizon).
Consider a Reissner-Nordstr\"om black hole whose the horizons are given by $r_\pm= M \pm \sqrt{M^2-Q^2}$ (in the units $c=G=4\pi \epsilon_0=1$). We assume $Q>0$, without loss of generality, for simplicity of notations. The black hole becomes extremal when $M=Q$. Clearly $r_\pm$ is no longer real when $Q>M$, which means that the singularity becomes naked. The most obvious and natural attempt to violate cosmic censorship is therefore to try to over-charge a black hole by throwing in a charge particle such that $(Q+e)/(M+m) > 1$. Similarly, one can try to overspin a rotating Kerr black hole to expose its ring singularity, by throwing in particles with the right amount of angular momentum.

However, it turns out that such a process is usually not easy, as shown by many works after Wald's landmark paper \cite{Wald0}. As a black hole gets closer to becoming extremal, it becomes harder to throw in particles that might otherwise destroy its horizon. This is due to the effective potential around the black hole that deflects the particles away. For the charged case, this can be appreciated rather straightforwardly (with the risk of over-simplifying): for a black hole with small amount of electrical charge, if we put an electron next to the black hole it will fall inwards due to the gravitational pull of the black hole. However, for a highly (negatively) charged black hole, the tendency is to repel the electron since gravity is so much weaker than the electromagnetic force. One could attempt to overcome this by throwing in the electron with large velocity. This energy, however, add to the mass of the black hole so that the overall effect is to prevent naked singularity formation. 
Thus, the common misconception that black hole is a dangerous place that would suck anything in is far from the truth -- of course black holes do not actively suck stuff in anymore than any object of the same mass does -- but furthermore, it is sometimes rather difficult to get into one! Even in those situations that such attempts appear to be successful at first \cite{0907.4146}, are not when backreaction of the particle's ``self-energy/force'' is properly taken into account \cite{0205005, 1008.5159, 1501.07330, Wald,1707.05862}. However, a recent work involving a magnetized black hole claimed to have violated the weak cosmic censorship, which is unlikely to be restored by considering self-force \cite{2003.12999}, though this should be checked more carefully. Test fields were shown to unable to violate the weak censorship conjecture \cite{0508011} as long as the null energy condition is satisfied \cite{1112.2382, 1601.06809, 2004.02902}.  The perturbation of Kerr-Newman black holes by neutrino fields violates the null energy condition, thus could lead to over-spinning of Kerr-Newman black holes \cite{1408.1735,1904.05185}. In \cite{2001.03106}, it was found that test scalar field cannot overspin a Kerr-Taub-NUT black hole, but near extremal one can be overspun with test particle. The authors suggested that the different results between test field and test particle indicate the time interval for particles crossing the black hole horizon is important.

In addition to dropping particles into black holes, one could also consider dropping a shell of matter into a black hole. While a sufficiently charged shell is not prevented from entering a black hole, a new, larger horizon forms -- and indeed does so before the charged shell crosses the original black hole horizon. This ensures that the weak cosmic censorship holds \cite{1302.6658}. 

On the other hand, partly motivated by theories with higher dimensions of spacetime such as string theory, and partly driven by pure mathematical curiosities, general relativity in dimensions above four has also been studied rather extensively. Various types of black holes are allowed in higher dimensions, including black string, black ring and black brane. We could ask if these black holes can become ``super-extremal'' to form naked singularities. Indeed, 
\cite{1003.4295} had looked into a variety of black hole geometries, from black holes in higher dimensions to black rings and showed that this particular way of destroying a black hole does not succeed. However,
many of these extended higher dimensional ``black objects'' are unstable against perturbations. Take a thin black string (and similarly a black ring), for example, which is an extended object in say, the $z$-direction (spacelike). That is, each constant $z$ slice is a Schwarzschild black hole. Under perturbation, an instability similar to the Rayleigh-Plateau instability of fluid mechanics arises \cite{0602017}, so that ``lumps'' and ``necks'' start to form along the $z$-direction, and each constant $z$ slice is no longer of the same size. 

The thinning eventually progresses until the string or ring pinches off at multiple places. The pinch-off is believed to occur in finite time, and when that happens the singularity will be exposed. Due to this phenomenon, called the Gregory-Laflamme instability, the general impression is that cosmic censorship is easier to be violated in spacetime dimensions above four (and so perhaps it is a good thing that we live in a 4-dimensional world). Indeed, a varieties of solutions describing naked singularity formation in gravitational collapse have been found in $6$ dimensions \cite{1509.07956}. This is consistent with the recent findings that in higher dimensions, when two black holes collide, \emph{less} energy is radiated in gravitational waves, and as a result it is also easier to create naked singularities when higher dimensional black holes collide \cite{1812.05017}. Formation of naked singularities in this manner becomes more efficient as the number of spacetime dimensions increases. See also \cite{1105.3331} and \cite{1909.02997}.

What we have discussed in this section until now, are about the \emph{weak} cosmic censorship. However,
the reason that cosmic censorship conjecture is again receiving a lot of attention recently, is partly due to two discoveries, on both the physical as well as mathematical front. On the physics side, it is the discovery that a highly charged black hole in de Sitter spacetime can violate the \emph{strong} cosmic censorship conjecture \cite{1711.10502, 1808.03631, 1811.08538, 1902.01865}. One can understand this in a somewhat hand-wavy manner. Strong cosmic censorship is safe if the Cauchy horizon can be destroyed by the blueshift instability. However, in a de Sitter universe, there is also a cosmological redshift that competes with the blueshift at the inner horizon. As a result the blueshifting of energy is not enough to de-stabilize the Cauchy horizon. This result is especially interesting since cosmological observations suggest that our Universe is asymptotically de Sitter. One might object that highly charged black holes are not physical in the actual Universe, but one can always imagine creating such a black hole artificially by dropping in charged matter. The point is that one does not have to over-charge the black hole, but merely making it charged enough. One might ask why we need the black hole to be sufficiently charged, since if cosmological redshift can overcome blueshift instability for a highly charged black hole, why it cannot do so for weakly charged holes? This is where details matter.

The actual characterization depends on the quantity $\Gamma \equiv \text{Im}(\omega_0)/\kappa_-$, where $\omega_0$ is the fundamental quasinormal resonant frequency that governs the decay rate of black hole perturbation $\psi(t\to \infty) \sim e^{-\text{Im}(\omega_0)t}$, while $\kappa_-$ is the surface gravity of the Cauchy horizon that governs the blueshifting effect that grows like $e^{\kappa_-v}$ where $v$ is the advanced null coordinate (see, e.g., \cite{1801.07261,1810.12128,1910.09564}). Clearly if $\kappa_-$ is much larger than $\text{Im}(\omega_0)$, the blueshift instability will win and cosmic censorship is safe. If the decay rate is large enough however, then it can weaken the blueshift effect and cause the inner horizon to be stable. It turns out that for the strong cosmic censorship to be valid, one requires the existence of at least one quasinormal resonant frequency that satisfies $\Gamma \leqslant 1/2$. We also see that highly charged black holes are in more danger of violating censorship since $\kappa_-$ is smaller (and tends to zero in the extremal limit). 

Indeed, starting in the 1990s, there have been other evidences that strong cosmic censorship can be violated in de Sitter spacetime \cite{9709025}. More recently, it was discovered that there may be topology-changing transitions for Reissner-Nordstr\"om de Sitter black holes and Gauss-Bonnet de Sitter black boles in higher dimensions that hint at violation of the weak cosmic censorship \cite{1909.02685}. 

On the more mathematical side, Dafermos and Luk \cite{1710.01722} showed that even the Cauchy horizon of an \emph{asymptotically flat} Kerr black hole (assuming that the exterior Kerr spacetime is indeed dynamically stable, which is still yet to be proved) is not as singular as previously expected from mass inflation mechanism. Instead it forms a ``weak null singularity'' \cite{1311.4970}, which allows the metric to be continuously extended beyond it, so strong cosmic censorship is violated in the strict sense. However, it cannot be made sense of even as a weak solution to the Einstein field equations, so it is physically unclear what happens to an observer that falls towards the inner horizon. If the Einstein field equations can no longer be used to describe an infalling observer beyond the inner horizon, can one really say that cosmic censorship has been violated? On the other hand, quantum effects might nevertheless destabilize the Cauchy horizon \cite{9411002}. See also \cite{0808.1709,1010.2585}.

Recently, I have looked into the whether Hawking radiation can lead to the formation of naked singularity. The idea is rather simple. Since the physical parameters of the black holes (such as its mass, charge, and angular momentum, i.e. the black hole hairs) can change their values under Hawking evaporation, one could ask, for example, in the case of Reissner-Nordstr\"om black hole, whether the charge-to-mass ratio $Q/M$ can increase towards extremality (or $a/M$ in the case of Kerr black hole). Davies recognized a long time ago that such a violation of the third law of black hole thermodynamics (since extremal Reissner-Nordstr\"om black hole has zero temperature) is equivalent to violating the weak cosmic censorship \cite{davies1977}. This is because a slight perturbation might cause an extremal black hole to become super-extremal (i.e. a naked singularity)\footnote{It is true that extremal Reissner-Nordstr\"om black hole saturates the BPS (Bogomol'nyi-Prasad-Sommerfield) bound -- and therefore minimizes energy at fixed charge -- and preserves some supersymmetry. However BPS solutions in a \emph{(super)gravitational theory} can be unstable \cite{1212.2557}, due to the lack of a local gravitational energy density. The situation is similar in classical general relativity: although there is a positive energy theorem, this alone does not guarantee even the (nonlinear) stability of Minkowski spacetime, let alone that of a black hole. Indeed, extremal Reissner-Nordstr\"om black hole suffers from classical, \emph{linear}, Aretakis instability \cite{Aretakis1, Aretakis2, Aretakis3, LMRT, MRT, ZWA} as well as unstable against pair creation via the charge channel (Schwinger mechanism); see discussion later in the text.}. In the case of asymptotically flat Reissner-Nordstr\"om black hole, extremality is never reached, despite the fact that the charge-to-mass ratio $Q/M$ can \emph{increase} as the black hole evaporates due to both Hawking effect and Schwinger effect. Hiscock and Weems \cite{HW} had shown that in the so-called mass dissipation regime, where initial $Q_0/M_0$ is relatively small, since charge loss is inefficient, the overall effect is to increase $Q/M$. Even though $Q/M$ can increase and come close to extremality, it eventually turns around and evolves towards the Schwarzschild limit (for relatively large initial $Q_0/M_0$, the black holes are in the charge dissipation regime; they simply discharge monotonically towards the Schwazschild limit). In other words, extremality is never reached, and the cosmic censorship is safe\footnote{The Hiscock and Weems model has quite a few assumptions. In particular, it is only valid for isolated large black holes (of typical astrophysical size), so one cannot study how small black holes evolve. In addition, the method cannot be trusted very close to extremality (since Schwinger effect becomes very efficient at $Q/M=1$ and beyond, and so by continuity one expects the model to start to break down shortly before $Q/M=1$). It is therefore a good thing that black holes stay \emph{away} from extremality under evaporation. The full picture of charged black hole evolution would require further, more detailed, investigations.}. The same holds in higher dimensions \cite{1911.11990}.

It is worth mentioning that in the gravitational collapse of a spherical shell of charged, massless scalar field, weak cosmic censorship is also respected. As one increases the initial charge-to-mass ratio, the final
black hole charge-to-mass ratio reaches a maximal value smaller than unity and then decreases \cite{0306078}. This is due to the increased electrostatic repulsion of the charged shell. Furthermore, in a recent study that models black hole physics with moving mirrors, Good found that when a null shell collapses to form an extremal Reissner-Nordstr\"om black hole, the formation of a naked singularity is prevented by emission of a finite amount of energy during the collapse.  The final extremal black hole cannot emit any more neutral particles (via non-thermal particle production) with enough energy to exceed the extremal limit \cite{2003.07016}. 

In the case of charged dilatonic ``GHS'' \cite{ghs, g, gm} black hole, which is a solution in low energy limit of certain string theories, I have shown \cite{1907.07490} that if we naively apply Hiscock and Weems model (with the same charge loss rate given by flat space quantum electrodynamics), then for a large enough mass, the GHS black hole can violate weak cosmic censorship. This means that the charge loss rate is not efficient enough near the singularity (which, for the GHS black hole, is a null singularity that coincides with its horizon, which shrinks to zero size at extremality). If we wish to \emph{impose} cosmic censorship, this means that we need to enhance charge loss rate. The simplest ansatz one can consider is of the form (since the exponential term can only involves dimensionless quantity, and it should probably depends on the charge of the black hole, hence the natural quantity is $Q/Q_0$, where $Q_0:=\hbar e/(\pi m^2)$ is the inverse of the Schwinger critical field, with $m, e$ denoting the mass and charge of an electron, respectively):
\begin{equation}\label{dQdtnew}
\frac{\d Q}{\d t} \approx -\frac{e^4}{2\pi^3\hbar m^2}\frac{Q^3}{r_+^3}\exp\left(-\frac{r_+^2}{Q_0Q} + \mathcal{C}\frac{Q}{Q_0}\right),
\end{equation}
where $\mathcal{C}>0$ is a constant that is absent in ordinary flat space QED.
By considering a near extremal GHS black hole, i.e., $Q=\sqrt{2} M - \varepsilon$ for some small $\varepsilon >0$, the exponential term can be expanded into series in $\varepsilon$:
\begin{flalign}
\exp\left(-\frac{r_+^2}{Q_0Q} + \mathcal{C}\frac{Q}{Q_0}\right) = &\exp\left(\frac{\sqrt{2}M (\mathcal{C}-2)}{Q_0}\right) \\ \notag &\times \left[1-\frac{2+\mathcal{C}}{Q_0}\varepsilon + O(\varepsilon^2)\right].
\end{flalign}
As long as $\mathcal{C} < 2$, the exponent is suppressed by the leading order term. The suppression effect increases with $M$, this is why $Q/M$ for sufficiently large $M$ increases steadily to extremality. That is to say: for any fixed value of $0 < \mathcal{C} < 2$, we can always choose $M$ sufficiently large so that there exist initial datum that lead to extremality under Hawking evaporation and Schwinger pair production. Imposing cosmic censorship therefore suggests that we choose $\mathcal{C}=2$. Remarkably, this result agrees perfectly with the direct calculation to obtain dilatonic correction to QED on the curved background of GHS black holes obtained by Shiraishi \cite{1305.2564}, which makes no reference to cosmic censorship. This complements previous study that GHS black holes can neither be overcharged nor be driven to extremality by interaction with test particles \cite{1812.06966}.

This result suggests that cosmic censorship can be used as a principle to deduce other physics (in this case, how dilaton coupling would modify QED). 
Furthermore, this result also suggests that Hawking radiation and other particle production processes are likely to respect cosmic censorship. To check if this is indeed the case, one has to check the evaporation of other nontrivial black hole solutions. The rotating case is vastly more complicated, of course.  An asymptotically flat Kerr black hole evaporating by the emission of \emph{only} scalar radiation will evolve towards a state with rotation parameter $a \approx 0.555 M$, i.e. if the black hole starts out with $a$ smaller than that, it will spin up towards said value; and if the initially angular momentum is large enough it will spin down towards said value. Adding higher spin particles, however, the black hole will spin down towards Schwarzschild limit unless there are many more species of massless scalars \cite{9710013, 9801044}. Thus, even in the rotating case, it seems that Hawking radiation will not bring a black hole towards the extremal state. In a more recent work \cite{1508.06685}, it was further argued that Hawking radiation can evolve a rotating black hole below the mass range $M \sim 10^{17}-10^{19}$g away from extremality even if a test field is sent in to overspin the black hole, provided that the test field is sent into the black hole from a distance sufficiently far.  

Since we have mentioned Hawking radiation, we should also mention at least two other contexts in which naked singularities might appear under Hawking evaporation. One possibility is at the \emph{very late time} at the end of the Hawking evaporation process. Whether it is a Schwarzschild black hole or a charged one, the final limit seems to be the same: a complete evaporation that leaves nothing behind. Or at least, according to conventional wisdom. I do not think it is obvious that a spacetime singularity can ``evaporate'' away. Might it be possible that at the very end of Hawking evaporation a naked singularity formed? It is plausible\footnote{Penrose wrote in \cite{penrose1999}: ``\emph{It is hard to avoid the conclusion that the endpoint of the Hawking evaporation of a black hole would be a naked singularity -- or at least something that one} [sic] \emph{a classical scale would closely resemble a naked singularity.}'' However, Penrose was more concerned with cosmic censorship in classical general relativity.}. If so, the end state might be a zero mass naked singularity. These objects may not be benign as they may produce a divergent flux of particles \cite{hiscock}.

However in view of the preceding conjecture that Hawking evaporation respects cosmic censorship, perhaps a backreaction to the spacetime (as the temperature becomes extremely high towards the end of Hawking evaporation) might modify the process so much so as to change the picture completely. Perhaps quantum gravitational effect prevents a complete evaporation (this is known as the black hole remnant scenario \cite{1412.8366}), and therefore saves cosmic censorship.

Another context in which naked singularity might appear concerns black hole firewall \cite{amps, apologia}, a putative high energy wall that replaces a smooth black hole horizon, which would completely obliterate any would-be explorer who attempts to fall into the black hole (see also, \cite{0907.1190}). The idea of firewall goes as follows. Suppose a black hole is formed from a pure state. By unitarity of quantum mechanics, the end result after Hawking evaporation should also be a pure state. The information that falls into a black hole is not lost, but rather encoded in the Hawking radiation via quantum entanglement, though in a highly scrambled form. Roughly speaking, the first half of the Hawking particles emitted when the black holes are ``young'' carry no information. The information only starts to ``leak out'' of the black hole as the late time Hawking evaporation gradually purifies the earlier radiation. (The transition between early and late time is called the Page time \cite{page1,page1b,page2}.) However, in order to do so, the late time radiation must be maximally entangled with the early radiation. Since quantum entanglement is ``monogamous'', this implies that the late time Hawking particles cannot, at the same time, be maximally entangled with their partner particles that fall into the black hole (some authors have doubted the maximal entanglement assumption, see, e.g., the recent work of Dai \cite{1908.01005}). This leads to a discontinuity of the near horizon quantum field, which in turn leads to a divergent in its Hamiltonian. Physically this means that there is a very high energy wall around the black hole, the firewall. (See \cite{1409.1231} for a good introduction on black holes and quantum information.)

There are many proposals from both the proponents and the opponents of the firewall idea. I belong to the latter camp, since firewalls seem like a drastic violation to our understanding that in general relativity, there is nothing peculiar at the horizon in the sense that for a large enough black hole the tidal force there is negligible (spacetime curvature is small).  My collaborator and I have argued that \emph{if} black hole firewalls exist, then they could (with some probabilities) become naked, even arbitrarily far away from the black hole. We call these ``naked firewalls'' \cite{1511.05695}. In other words, an observer light years away from any black hole might suddenly encounter one such ``naked firewall'' and be destroyed. This, we argue, cannot be ``conservative'' as the firewall proposal is supposed to be. This is our \emph{reductio ad absurdum} argument against the existence of firewalls. Needless to say, ``conservative'' is in the eyes of the beholders, so we do not expect this finding to dissuade a hardcore believer in firewalls. 

The point is that our naked firewall is in many ways similar to a naked singularity. In fact, Susskind once proposed, based on the idea that spacetime region is emergent out of quantum entanglement \cite{1005.3035, 1204.1330, 1206.1323}, that the firewall \emph{is} the singularity of a black hole that gradually migrates ``outward'' as the black hole evaporates \cite{1208.3445}. That is to say, as maximal entanglement between the black hole interior and exterior slowly vanishes, the (temporal) distance between the horizon and the singularity becomes smaller and eventually there is no spacetime between the horizon and the singularity, as the two eventually coincides. This is also similar to the extremal GHS null singularity (and perhaps, also similar to a more realistic extremal charged black hole even in general relativity \cite{1005.2999}).

\section{Curiosities and Prospects}\label{sec4}
\begin{quote}
\emph{As regards this little key, it is the key to the
small room at the end of the long passage on
the lower floor. You may open everything,
you may go everywhere, but I forbid you to
enter this little room. And I forbid you so
seriously that if you were indeed to open the
door, I should be so angry that I might do
anything.}
-- ``La Barbe Bleue'', Charles Perrault
\end{quote}

We have seen that under certain circumstances, naked singularities do form, and in some other situations, predictability is lost inside black holes, although exterior observers are ignorant about this. Should we be worried about either situation? In the case of naked singularities that form when a black string or a black ring pinches off, the pinching involves only Planck size area, and so in the process, likely only emits a few Planck energy quanta in some Planck time, before they become disconnected black holes. Therefore the threat to classical predictability is very small -- evolution up to, and after the pinching, can be described purely by classical general relativity. This is also the case for the naked singularity formation under black hole collision in higher dimensions. 
So, to quote \cite{1812.05017}, ``even if cosmic censorship is violated, its spirit remains unchallenged''. 
The danger is more serious in the case of, say, GHS extremal black hole (of zero size), since its curvature becomes unbounded towards the extremal limit. In that case the situation is the opposite, even if the censorship is not violated in substance, it is violated in spirit. Under some perturbations, a \emph{bona fide} violation could happen. It is comforting to know that Hawking radiation and Schwinger effect do not allow extremality to be reached and thus prevented this danger from being realized (at least in the regime where the method we discussed above is trustable). 

Regardless, the attitude towards violation of cosmic censorship has changed in the recent years. These are no longer treated as abominations that one should \emph{definitely} get rid of in the theory of general relativity (or other theories of gravity). Rather, since singularities should carry information about quantum gravity, any signal from them should be welcomed. They are windows of opportunities into the as yet unreachable realm of quantum gravity\footnote{An analogy from fiction: Frankenstein's monster should be treated as an opportunity to further understand the mystery of life, and furthermore, the meaning to be alive, rather than a fearsome creature.}. Exactly what are the observational signatures of naked singularities, shall they exist, have also been discussed in the literature, ranging from gravitational lensing to accretion disk, photon orbits, perihelion precession, and shadows \cite{VE, 0806.3289, 1206.3077, 1801.00867, 1802.08060,1811.02648,1909.10322,2003.06810}. Unfortunately thus far astronomical observations have yet to locate any real evidence for their existence.  

One current main objective should be to study under which physical processes can cosmic censorships be violated. This will at least help us to understand how to properly refine (define?) the conjecture: how should the word ``generic'' be understood? Should we only be concerned with dynamically formation processeses \cite{1910.09564}? 

Another direction of research is to understand whether there are further connections between cosmic censorship conjectures and other conjectures in gravitational and high energy physics. One such link with the weak gravity conjecture has been found. The weak gravity conjecture \cite{0601001} is essentially the statement that gravity is always the weakest force. To be more specific, it states that the lightest charge particle with mass $m$ and charge $q$ in any $\text{U}(1)$ gauge theory that admits an ultraviolet embedding into a consistent theory of quantum gravity should satisfy the bound
\begin{equation}
\frac{q}{\sqrt{\hbar}} \gtrsim \frac{m}{m_\text{Pl}},
\end{equation}
in which the coupling constant of the $\text{U}(1)$ force has been absorbed into the normalization of the charge $q$. 
In anti-de Sitter spacetime, it turns out that one could introduce a sufficiently strong electric field that grows without bound, which in turn produces arbitrarily large spacetime curvature that is visible to infinity \cite{1604.06465, 1702.05490}. A true naked singularity never forms, but again, the fact that arbitrarily large curvature is allowed means that the weak cosmic censorship is already violated. 
However, if weak gravity conjecture holds, then the bulk \emph{must} allow for charged matter, which was shown to destabilize the electric field (essentially via Schwinger effect that produces charged particles) and thereby preventing the violation of cosmic censorship \cite{1709.07880}. Remarkably, the minimum value of charge required to preserve cosmic censorship appears to agree precisely with that proposed by the weak gravity conjecture. Further evidence for such a connection was found in \cite{1901.11096}, in the presence of two Maxwell fields, and in the presence of a dilaton coupling. As mentioned by Horowitz and Santos in \cite{1901.11096}, this relation between cosmic censorship and weak gravity is intriguing since even though the current attitude towards cosmic censorship is that its violation could provide some hints about quantum gravity, a conjecture about quantum gravity is now \emph{preserving} cosmic censorship. 

This might imply that other violations of the cosmic censorship can be prevented if quantum gravitational effect is taken into account. Indeed, even at the semi-classical level, it has been argued that backreaction could lead to formation of horizon around a singularity, thereby endowing it with a ``quantum dress'' \cite{1605.06078}. See also \cite{1686412} and \cite{1908.07402}. 

In general, there is a need to understand singularities in quantum gravity theory such as string theory, as arbitrarily high curvature is not necessarily a bad thing either \cite{0002160}. In addition, quantum gravity might also modify the cosmic censorship bound for the black hole parameters, i.e. what appears as naked singularities in classical general relativity might cease to be one in quantum gravity. Such a scenario has been investigated in string theory \cite{0706.2873, 1707.07242}.  If this is the case however, it is not so clear why quantum gravity would modify the censorship bound, which actually concerns the existence of the event horizon, \emph{not} the singularity -- the latter becomes naked when the horizon disappears, it is only the consequence of the violation of the censorship bound. At the singularity one expects quantum gravitational effect to become more important, but certainly at the horizon where curvature can be arbitrarily small, one should not expect a large quantum correction?

The relation between cosmic censorship and weak gravity conjecture in de Sitter spacetime has also been investigated recently by considering charged particle production of black holes via Hawking and Schwinger effects \cite{1910.01648}. Furthermore, there is evidence that naked singularities are related to causality, at least in the case of anti-de Sitter spacetimes \cite{1903.11806}.

In addition to string theory, one should also pay attention to other approaches of quantum gravity that might offer some insights into spacetime singularities and their possible resolutions.
Black holes in asymptotically safe gravity is one such scenario \cite{1503.06472}. The basic idea is to allow Newton's gravitational constant to run under renormalization group flow, an idea borrowed from quantum field theory, to improve the quantum behavior of general relativity. Under such scenarios, it is possible to either resolve classical singularities under some conditions \cite{1808.03472, 1904.04845}, or at least ``soften'' it \cite{1610.05299}. In \cite{0810.0079}, Hod also argued that, as once quantum mechanics is taken into account, absorption of fermion particles by a spinning black hole will not result in a naked singularity, cosmic censorship is inherently a quantum phenomenon despite it was first formulated in classical general relativity. However, this viewpoint was challenged in \cite{0905.1077}, which argued that the ``cosmic censor" may be oblivious to processes involving quantum effects.

In loop quantum gravity and its inspired models, intriguing scenarios have been explored. For example, black hole singularities might be avoidable if quantum gravitational effect causes a ``bounce'' \cite{1401.6562, 1404.5821, 1804.02821}, or if black holes tunnel into white holes \cite{1407.0989, 1801.03027, 1802.04264, 1905.07251,1911.12646,2004.13061}. Conditions under which such processes can indeed happen, taking into account stability issues had been investigated in \cite{1511.00633}. It was found therein that transitions with long characteristic time scales are pathologically unstable, which prevents transition into white holes. Geometries with short characteristic time scales are, however, robust against perturbations. Another interesting possibility pointed out by Bojowald is that the core of the black hole becomes Euclidean (that is, with four spatial dimensions) due to quantum gravity effect \cite{1409.3157}. Such a signature change causes the spacetime to become indeterministic (there is a Cauchy horizon), and although there is no true singularity, the same problem one has with naked singularities concerning the loss of predictability occurs in this scenario. Nevertheless,  it is not clear if such a Euclidean core is a real fundamental effect or simply an instability in the equation of motion \cite{1607.07589}. It also remains to see how the ``Euclideanization'' effect might affect the black hole to white hole transition mentioned above \cite{1607.07589}. In a recent review, Ashtekar still holds the point of view that singularity should be resolved in loop quantum gravity \cite{2001.08833}.

Lastly, I will let my imagination loose and make some wild speculations to end this review. 
For a long time, the attitude is to \emph{not} speak about the singularity at all, apart from the relatively small community that continues to investigate cosmic censorship. This, in some way, parallels the so-called ``shut up and calculate'' mindset of many theoretical physicists concerning the foundation of quantum mechanics. That is, the practical view that interpretations of quantum mechanics -- of what is the underlying nature of quantum mechanical laws -- is not important as long as we can make physical predictions. 
Since we do not know how to describe singularities, it does make sense to ignore it and proceed to understand what we \emph{can} study. After all, general relativity is rich enough to provide us with century worth of discoveries even if we stay far away from the mysterious nature of spacetime singularities. 

However, it is perhaps not wise to sweep singularities under the carpet entirely, hoping that the magic of quantum gravity will resolve it, for it may not be resolvable even then. Or perhaps by actually taking singularities seriously, we might learn a thing or two about the true nature of the -- as yet elusive  -- quantum gravity.
Likewise, the interpretation of quantum mechanics is said to be a more philosophical question than a physical one, because as long as they give the same predictions for experiments, then one cannot distinguish between them (though, see, e.g., \cite{1403.1188}). 
However, it is also possible that some interpretations are easier than the others to unify with general relativity, because different formulations that are equivalent at this stage, could lead to \emph{different} theories once subsequently generalized. As Feynman put it \cite{feyn}: ``In other words, although they are identical before they are changed, there are certain ways of changing one which look natural, which don't look natural in the other. Therefore, psychologically, we must keep all the theories in our head.'' 
Perhaps eventually different interpretations will give different testable predictions about quantum gravity. After all, singularities are where we expect both quantum physics and gravity to play equally important roles. (Similarly, Hsu has argued that an experiment capable of illuminating the information paradox of black holes must necessarily be able to detect or manipulate macroscopic superpositions \cite{1003.5382}).
Might it be that what we tend to ignore today -- singularities and interpretation of quantum mechanics -- are closely related? 

Is the cosmic censor a ``law of physics''? If so, what is its range of validity? When does it apply? What is its true underlying mechanism? Are there more than one censors at work? What can it tell us about quantum gravity? There are still many open questions to be explored.

\begin{acknowledgments}
The author thanks the National Natural Science Foundation of China (No.11922508, No.11705162) and the Natural Science Foundation of Jiangsu Province (No.BK20170479) for funding support. He also thanks Brett McInnes for related discussions.
\end{acknowledgments}

\end{document}